\documentclass[aps,showpacs,tightenlines,amssymb,twocolumn]{revtex4}
\usepackage{epsfig} \usepackage{bm}
\newcommand{\be}{\begin{equation}}\newcommand{\ee}{\end{equation}}
\newcommand{\bea}{\begin{eqnarray}}\newcommand{\eea}{\end{eqnarray}}
\def\littlespace{$\,$}
\def\unit#1{\hbox{\littlespace #1}} 

\def\halftext{\textstyle{\frac12}} \def\htx{{\halftext}}
\def\quartertext{{\textstyle{\frac14}}}
\def\Eqref#1{Eq.~(\ref{#1})}
\def\oms{\omega_s}
\def\omone{\omega_1}
\def\quasisection#1{\medskip\noindent\textit{#1.}~~}

\newcommand{\labeld}[1]{ }

\begin{document}
\title{Stability of quantum breathers}
\author{L. S. Schulman\footnote{Email: schulman@clarkson.edu}}
\affiliation{Physics Department, Clarkson University, Potsdam, New York 13699-5820, USA}
\author{D. Tolkunov
}
\affiliation{Physics Department, Clarkson University, Potsdam, New York 13699-5820, USA}
\author{E. Mihokova
}
\affiliation{Institute of Physics, Academy of Sciences of the Czech Republic, Cukrovarnick\'a~10, 162~53~Prague~6, Czech~Republic}
\date{\today}
\begin{abstract}
Using two methods we show that a quantized discrete breather in a 1-D lattice is stable. One method uses path integrals and compares correlations for a (linear) local mode with those of the quantum breather. The other takes a local mode as the zeroth order system relative to which numerical, cutoff-insensitive diagonalization of the Hamiltonian is performed.
\end{abstract}

\pacs{31.15.Gy, 31.70.Hq, 05.45.-a, 63.20.Pw.}

\maketitle

Discrete breathers \cite{note:names} are nondispersive classical excitations that are known to be significant in a variety of natural systems. Recent reviews \cite{campbell,fleurovchaos} give examples and references. The quantum theory of (discrete) breathers does not yield easily to numerical simulation and is not as fully developed.

Over the past decade there has developed a body of literature \cite{hiszall} in which it is claimed that unlike classical breathers, the quantum version cannot in principle be stable. This conclusion is reached by considering the effect on the phonon field of the rapidly oscillating particle at the core of the breather. Lifetimes have been calculated, and for alkali halides are about 10\unit ns. Other work, also treating the breather as an external classical force, \cite{flachfleurovradiation}, challenges this estimate and suggests stability in the limit of large systems.

A fully quantum result \cite{wang} found eigenstates of a model Hamiltonian with a dropoff in correlation function that is strongly suggestive of breathers, although translational invariance prevents complete localization. As recognized in \cite{wang} (and below), such approaches require cutoffs which may affect the conclusion.

In this article we make a two-pronged attack. We do a numerical diagonalization, similar to that of \cite{wang}, incorporating methods to reduce cutoff effects and to provide direct evidence of localization. In addition we do a path integral calculation, treating the breather fully quantum mechanically, similar to Feynman's polaron \cite{feynmanpibook}. The approximations are different: in one case a cutoff, in the other semiclassical asymptotics.

The Hamiltonian in \cite{wang} is
\be
H=\textstyle{\sum_{k=0}^N} \left\{\htx p_k^2 +\htx\omega_s^2 x_k^2
             +\htx \omega_0^2(\Delta x)_k^2
                           +\quartertext\lambda x_k^4\right\}
\,,
\label{eq:H}
\ee
\labeld{eq:H}
with $x_0\equiv x_{N+1}$, $(\Delta x)_k\equiv x_k-x_{k+1}$, and $x_k\in{\mathbb R}$. We also consider interactions of the form $\sum(\Delta x)_k^4$, which resembles our own model \cite{confine}. An important simplification is to remove all nonlinear interactions except those of particle-0, the one with large amplitude motion. This strategy has been adopted by other authors and we checked whether this affects the general properties of the classical breather. It does not \cite{bigstabbr}, a consequence of the remarkable breather property, namely that a central atom vibrates strongly, while its neighbors hardly move (so that nonlinear forces are negligible for them). Claims that the breather decays are based on this same approximation.

We emphasize the reasoning leading from a translationally invariant breather to what is effectively a nonlinear local mode. Although quantum tunneling requires that the breathers form a band, information on the classical breather corresponding to our excitation---including in the presence of moderate noise---indicates that tunneling is small, allowing us to drop translational invariance. This perspective has been adopted by others as well \cite{hiszall, flachfleurovradiation}. Having done this, we go beyond previous work by quantizing the breather atom itself. In other approaches one has \textit{de facto} a local mode, interacting with the lattice. We retain the nonlinear interactions (implicitly), while also quantizing the principal breather atom.

In both our approaches we make use of a \emph{linear local mode}. If in (the modified) \Eqref{eq:H}, one replaces $\lambda x_0^4/4$ by $\omega_1^2 x_0^2/2$ with large $\omega_1$, the resulting system is well understood both classically and quantum mechanically. You have both classical confinement and quantum stability. Our use of the local mode parallels others' treatment of the quantum breather as a phonon field in the presence of rapid classical oscillations. The impact of the breather \emph{cannot} be considered small. They treat that impact by means of a classical oscillating field. We use the quantum local mode. For the path integral, that use is the comparison with local mode correlation functions. For the diagonalization scheme we perturb around that mode.

\quasisection{Path integral approach}
The path integral allows the elimination of quadratic degrees of freedom at the expense of introducing nonlocal self coupling. Dropping quartic terms for all but $x_0$, the Lagrangian is
\bea
{\cal L}
  &=& \textstyle{\sum_{n=0}^N}
     \left\{ \dot x_n^2  -\omega_0^2 (\Delta x)_n  -\omega_s^2x_n^2 \right\}/2
     -\lambda x_{0}^4/4                      \nonumber \\
     &&\qquad  -\mu x_0(x_m+x_{N+1-m})\,.
\eea
The fictitious coupling, $\mu x_0(x_m+x_{N+1-m})$, allows study of localization for $m$ far from~0. The derivative (at $\mu=0$) of the propagator provides a (0-$m$)-correlation. By standard methods \cite{feynmanpibook, lsspibook, weiss} we path integrate degrees of freedom 1--$N$, which form a \emph{chain} with known normal modes. These modes see a forcing term from the so-far-unintegrated $x_0(t)$. The propagator, $G$, is a function of the endpoints of \emph{all} $x_k$. One takes the matrix element of $G$ in the chain ground state (\`a la \cite{feynmanpibook}) and divides by corresponding free chain matrix elements. The result is
\def\Seff{S_{\hbox{\tiny eff}}}
\be
{\cal G}\left( q^f,T;q^i,0\right) =\int {\cal D}q
              e^{\frac i\hbar \left( S_0+\Seff\right) },
\ee
where $q\equiv x_0$, the action $S_0$ arises from the original $\cal L$ sans chain terms, and $\Seff$ comes from the integration over chain degrees of freedom. Specifically
\be
S_0=\int dt\left\{ \dot q^2/2-\left(\omega _s^2/2+\omega _0^2\right) q^2-\lambda q^4/4\right\} \,,
\ee
\be
\Seff=\int_0^T\int_0^T \,dt\,ds\, K(|t-s|)q(t)q(s) \,,
\ee
with
\be
K(u)=\sum_{n=1,3,\dots}^N \tau_n^2
          \frac{\cos\Omega_n\left(\frac T2-u\right)}
                      {\Omega_n\sin\left(\frac{\Omega_n T}2\right)}  \,,
\ee
\be
\tau_n^2\equiv \frac2{N+1}\left[\omega_0^2\sin\frac{\pi n}{N+1}+\mu
\sin\frac{\pi nm}{N+1}\right]^2  \,,
\ee
and $\Omega_n^2=\omega_s^2+4\omega _0^2\sin ^2\left[{n\pi}/{2\left(N+1\right)}\right]$
is the spectrum of the chain.

We study $\cal G$ in the stationary phase approximation, that is we evaluate $S_0+\Seff$ along extremal ``classical paths.'' In principle one can get the ground state energy by going to large imaginary times, for which ${\cal G}(q,-iT;q)\sim |\phi(q)|^2 \exp(-T E_0)$. In practice $S\to\,$const and $\partial^2S/\partial q^i\partial q^f$ gives this exponentially small quantity (as for the harmonic oscillator). Such precision was not numerically possible; in any case, knowing the energy does not establish localization. Note that $\phi(q)$ is the overlap of the ground state of the $(N+1)$-atom ring with that of the $N$-atom chain.

To study localization we compare $\cal G$ with the propagator of a corresponding \emph{local mode}, i.e., a system in which $\lambda q^4/4$ is replaced by $\omega_1^2q^2/2$ for appropriate $\omega_1$ ($\sim\sqrt\lambda$). To derive ${\cal G}_{\hbox{\tiny local}}$
we redo the process described above. $K$ is the same. For this system we know that as $\omega_1$ increases, the vibrations are localized near~0. How can that be seen in ${\cal G}_{\hbox{\tiny local}}$ or $S$? The method is to go to imaginary time and vary $\mu$ (the coupling to atom \#$m$ $\approx N/3$).

Before presenting results we comment on numerics. The extremal solves a two-time boundary value problem with a nonlocal interaction. In the linear case we discretize time, so that $d^2/dt^2$ becomes a difference matrix, nearly diagonal, but the $K$-matrix is spread out. Call the resulting operator $B=[d^2/dt^2]+\dots$. Take $q$ to be a column vector, $q_j\leftrightarrow q(t_j$). For $q(0)=a$, $q(T)=b$, and $\epsilon=t_{k+1}-t_k$, the extremal is $q=B^{-1}q^{(0)}$, where $q^{(0)}_j=-(a,0,\dots,0,b)/\epsilon^2$ (cf.\ \cite{diffdiff}). Note that $S=q\dot q/2|_0^T$. To deal with computer limits on $\epsilon$, we calculated $S$ for the smallest practical $\epsilon$ and for \emph{larger} values. $S$ as a function of $\epsilon$ was then extrapolated to $\epsilon=0$. This technique was validated for the demanding energy calculation, involving the exponentially small $\partial^2S/\partial a\partial b$. Here the answer is known from the classical chain and ring frequencies. ($E_0$ also requires a $T\to\infty$ extrapolation.) The function $\phi$, a Gaussian whose spread can be calculated by standard methods, provided another check. Numerically that spread can be extracted by calculating $S$ with varying endpoints and fitting the slope of $\sqrt S$. This too confirmed the method.

\begin{figure}
\includegraphics[height=.25\textheight,width=.33\textwidth]{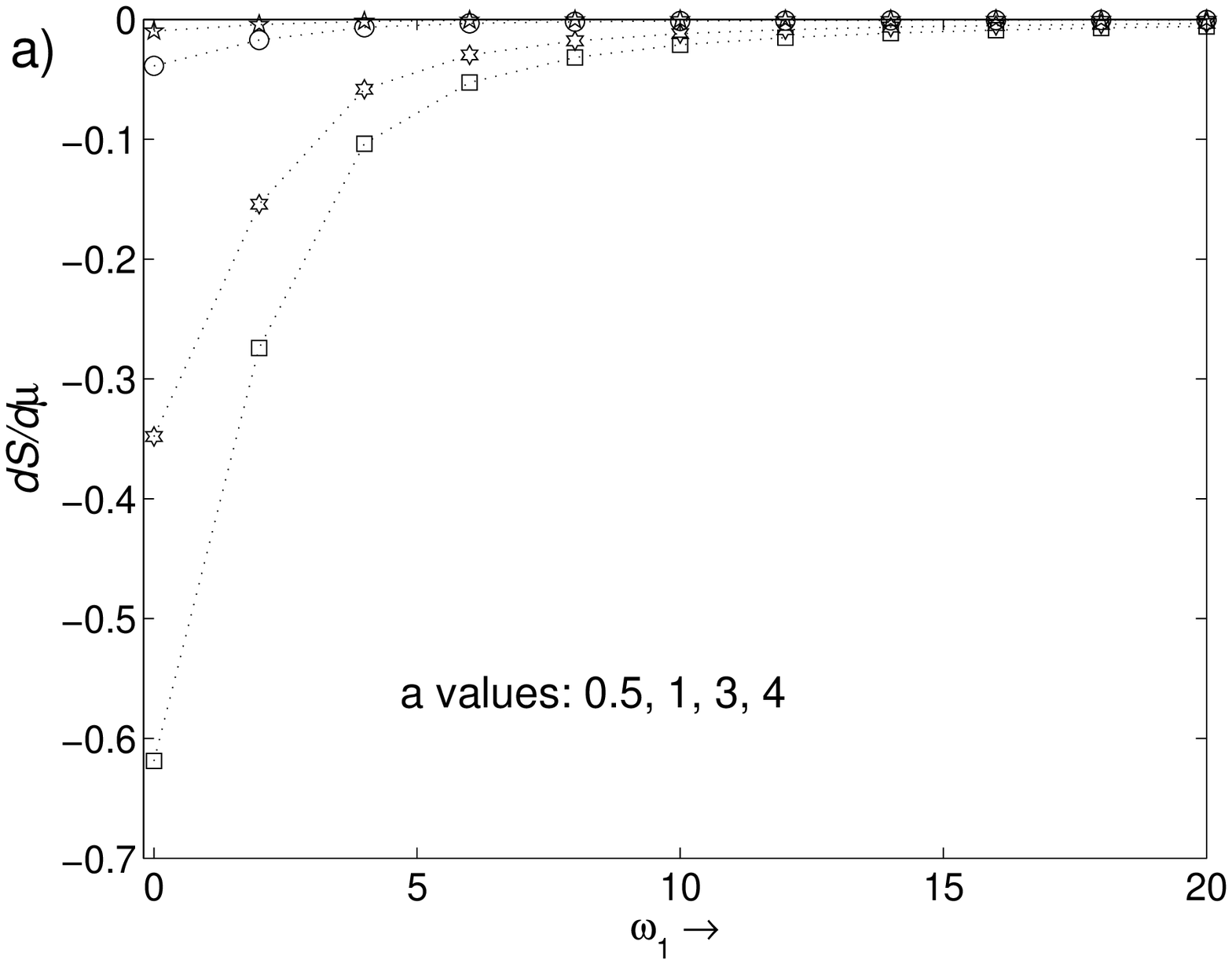}
\includegraphics[height=.25\textheight,width=.33\textwidth]{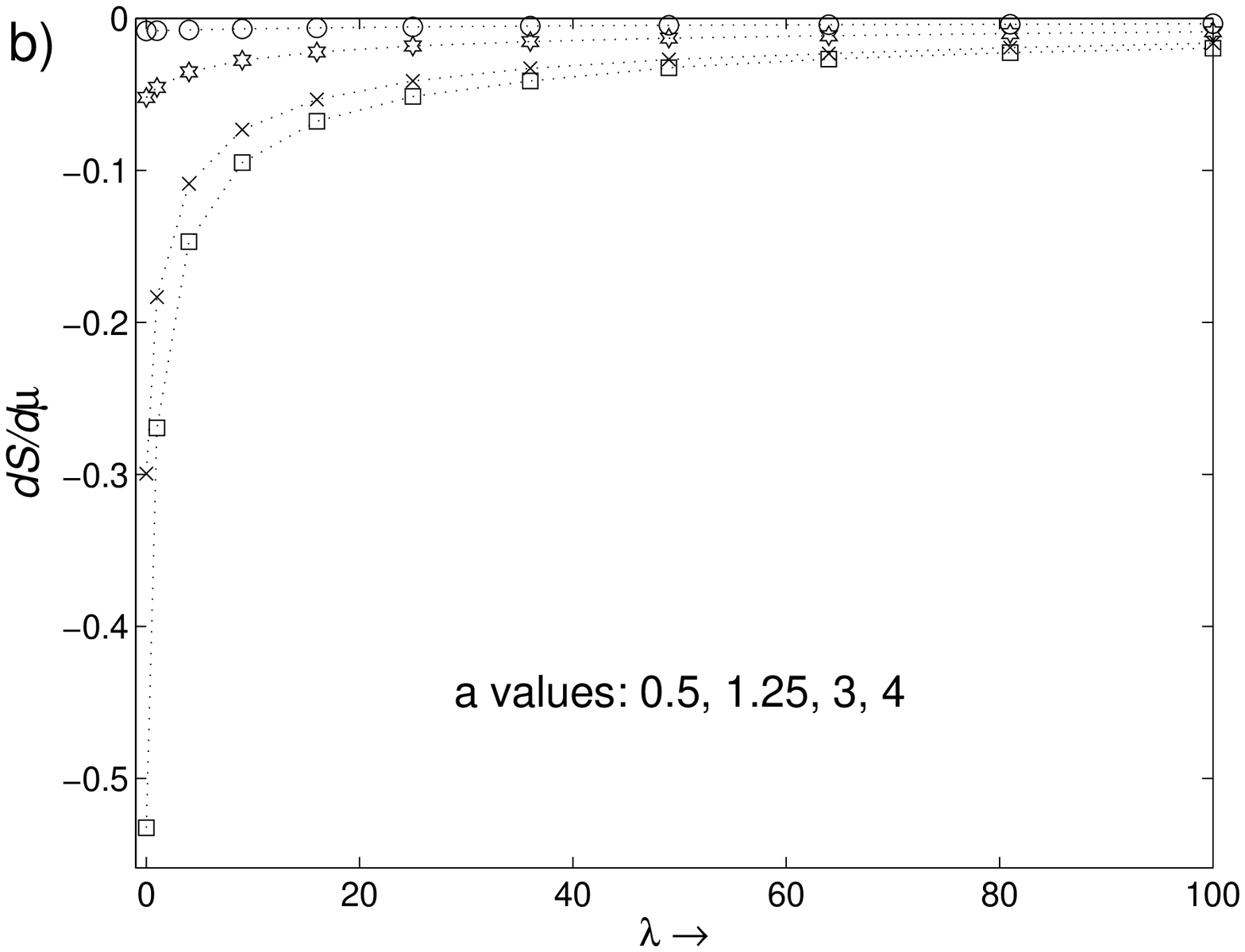}%
\caption{$dS/d\mu$ (essentially a correlation function) has markedly different behavior for large and small endpoint (``$a$'') values for $q$. Both for a linear local mode (as a function of $\omega_1$) and for the breather (function of $\lambda$), a large value of $a$ demands relatively large correlations when neither local mode nor breather is present ($\omega_1\sim\lambda\sim0$), but that correlation is wiped out for large parameter, for which the breather or local mode is effectively decoupled from the rest of the ring.}
\label{fig:vanishcorr}
\end{figure}

With $q^4$ the nonlinear, nonlocal two-time boundary value problem cannot be solved by matrix inversion. Our method was a variation of \cite{feynmanpibook}, defining a subsidiary variable $z(t)\equiv\int_0^T \widetilde K_1(|t-s|)q(s)ds$, where the subscript on $\widetilde K_1$ indicates that $\widetilde K$ ($=iK(-iu)$ with $T\to-iT$) is replaced by a single ``cosh'' with optimized parameters, in fact an excellent approximation. This allowed a local boundary-value numerical technique, with a self-consistency demand on $z(0)$. Because of the approximation we further attempted small variations to lower the true action, but they resulted in essentially no change.

Fig.\ \ref{fig:vanishcorr} shows the results of changing $\mu$ for the local mode (a) and for the breather (b). Localization is deduced from the variation of $dS/d\mu$ as a function of both $a$ ($=q^f\!=q^i$) and the coupling, $\omega_1$ or $\lambda$. All runs are for $T=2$, which (given the energy scale) allows many states besides the ground state to survive in a spectral sum for $\cal G$. Therefore a large $a$ boundary condition selects for the excited states, in particular for a localized excited state (breather or local mode) if there is one. When the coupling is small, there is no localized state; forcing $q$ to be large forces \emph{all} atoms on the ring to depart from their usual positions. Hence the large magnitude of $dS/d\mu$ on the left of both figures---for the largest $a$ values. (Small $a$ has almost no effect even for small coupling.) \emph{For large coupling however, there \emph{is} a localized state: both for the linear ($\omega_1$) and nonlinear ($\lambda$) systems, forcing $q$ to depart from 0 has almost no effect on a distant atom}. As further checks, in \cite{bigstabbr} we show that $dS/d\mu\to0$ for $\lambda\to\infty$, and that for both linear and nonlinear cases $dS/d\mu$ has essentially the same behavior as $T$ grows \cite{note:exponentialwavefunction}.

\quasisection{Diagonalizing a truncated Hamiltonian}
The Hamiltonian (\ref{eq:H}) is made finite-dimensional by using a phonon basis and imposing a cutoff on the level of phonon excitation. A low cutoff is needed because of the proliferation of dimensions in the (implicit) tensor product of phonon operators. The same problem was faced in \cite{wang} and we alleviate it by the following strategy. Instead of perturbing around free modes, we use modes of a fictitious (linear) system with a local mode at site-0. This reduces the amplitudes of other modes in the quantum breather. \Eqref{eq:H}, with nonlinearity confined to $x_0$, is written
\be
H=\sum_{k=0}^N \left\{\htx p_k^2 +\htx\omega_s^2 x_k^2
             +\htx \omega_0^2(\Delta x)_k^2\right\}
             +\htx \omega_1^2x_0^2
             + V_I
\,,
\label{eq:HLocal}
\ee
\labeld{eq:HLocal}
with $V_I\equiv \lambda x_0^4/4  -\omega_1^2x_0^2/2$ the perturbation. A variation on \Eqref{eq:H} uses a nonlinear \emph{coupling}, that is, terms $\lambda(\Delta x)_k^4/4$. Here too, we drop nonlinear terms except those (two) that include $x_0$, and add and subtract corresponding quadratics. As for (\ref{eq:H}), the classical dynamics is substantially the same.

We outline the calculation: The classical \emph{local} mode problem is solved, giving rise to (truncated) creation and annihilation operators in terms of which the perturbation is expressed. The operator $\hat x_0 = \sum(a_\ell+{a_\ell}^\dagger) {u_{\ell}}(0)/\sqrt{2\Omega_\ell}$,
where $u_\ell$ is the $\ell$-th mode of the system and $\ell=0$ is the local mode. The key to $\omega_1$'s effectiveness lies in $\{u_\ell(0)\}$. For $\omega_1=0$ these are all of about the same magnitude (for $N=20$ they are $\sim0.3$ for those not zero by symmetry). But with $\omega_1=3$, $u_0(0)\sim0.988$, while the (nonzero) others average $\sim0.04$. This severely reduces the amplitudes of phonons other than the local mode in the eigenstates of the full Hamiltonian.

With $\lambda=8$, $\omega_0=1$, and $\oms=1$ (used in \cite{wang}), and for $\omone=2.5$ the true eigenfunction having greatest overlap with the first excited state of the local mode is shown in Table \ref{tab:state}. Clearly the local mode dominates. The next largest component is the thrice excited local mode, which is merely a shape adjustment. Other modes barely make the $10^{-3}$ level. Note that the highest excitation level for \emph{other} phonons is 3, indicating that a cutoff of 6 is safe. In fact, even to probability $10^{-8}$ there is no excitation higher than 3 except for the local mode.

\begin{table}
\caption{Principal components of the breather state. The first 4 columns refer to the 4 symmetric phonons in a 6-atom ring. ($N$ atoms $\Rightarrow[N/2]+1$ symmetric modes.) Row~1: frequencies (local mode is highest). Subsequent rows: number-operator values. Fifth column: norm squared of the mode ($\log_{10}$ in parentheses). Cutoffs: local mode 13, others 6. For this state first order perturbation theory is good to 0.3\%. The last column reports the same calculation with cutoff~8.}
\label{tab:state}
\begin{tabular}{|c|c|c|c||c|c|}
\colrule
0.749  &   0.987  &    1.19 &     2.04 & Prob.\ (6) & Prob.\ (8) \\ \colrule
\colrule
0 & 0 & 0 & 1 & 0.9947    &  0.9952    \\   \colrule
0 & 0 & 0 & 3 & 3.315 (-3)&~3.321 (-3) \\   \colrule
0 & 1 & 0 & 0 & 1.122 (-3)& 0.848 (-3) \\   \colrule
0 & 0 & 1 & 0 & 5.748 (-4)& 4.345  (-4) \\   \colrule
1 & 0 & 0 & 0 & 1.545 (-4)& 1.167 (-4) \\   \colrule
0 & 1 & 0 & 2 & 1.004 (-4)& 0.754 (-4) \\   \colrule
1 & 0 & 0 & 2 & 2.866 (-5)& 2.152 (-5) \\   \colrule
0 & 0 & 1 & 2 & 2.303 (-5)& 1.728 (-5) \\   \colrule
0 & 0 & 0 & 5 & 3.670 (-6)& 3.627 (-6) \\   \colrule
\colrule
\end{tabular}
\end{table}

As a check of cutoff sensitivity we repeated this calculation with a cutoff of 8 (but the local mode still at 13). The results are in the last column of Table \ref{tab:state}. There is little sensitivity to the change---not only in the probabilities but in the composition of the state. Increasing ring size and reducing the cutoff preserves the pattern. For a ring of size 8 and a cutoff of 5 (10 for the local mode), the state was again dominated (Prob.\ 0.992) by the local mode, with the next contributor an excited local mode. Going to yet larger rings (size 10, and lower cutoff) preserves the pattern, with the first non-local mode phonon contributing at probability level 10$^{-4}$. Other breather states yield the same result: accuracy of perturbation theory, insensitivity to cutoff change, and eigenfunction domination by local modes.

For nearest-neighbor nonlinearity the story is the same. Systematic study of a size-8 ring with cutoffs of 6, 8 and 12, showed that the local mode dominated the Hamiltonian's ground state.

The message of this dominance is that the eigenstates of the Hamiltonian are well-localized, which is to say the quantized breather states are stable.

\quasisection{Quantum time dependence}
Although we have shown the breather to be dominated by the local mode, the point has been raised that by Fermi's Golden Rule \emph{any} coupling to a continuum implies instability. This is not true. Some years ago it was found that quantum systems with ample continuum coupling nevertheless could survive indefinitely \cite{limited}. This departure from ordinary decay depended on smoothness and threshold properties of the coupling. A related matter concerns normalizable bound states in the continuum~\cite{simon}.

\begin{figure}
\centerline{\includegraphics[height=.25\textheight,width=.4\textwidth]{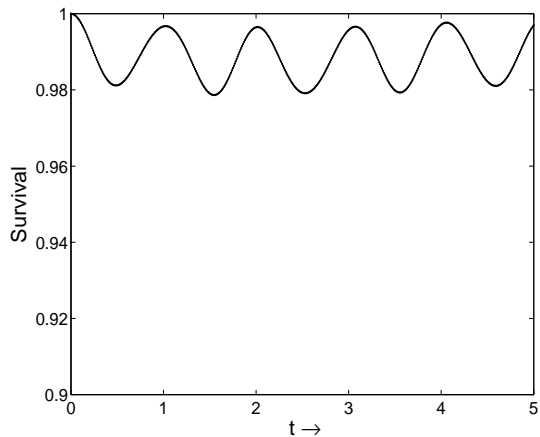}}%
\caption{Survival as a function of time; the initial state is the singly excited local mode.}
\label{fig:decay}
\end{figure}

In Fig.\ \ref{fig:decay} we show $|\langle\psi_0|\exp(-iHt/\hbar)|\psi_0\rangle|^2$ for $\psi_0$ a local mode phonon and $H$ the full Hamiltonian. As in \cite{limited}, initial decay is followed by stabilization bounded away from zero. In \cite{limited} the amplitude was typically a few tens of percent, but here, consistent with Table \ref{tab:state}, the amplitude stays near one. Despite the many connecting states, after a short time amplitude ceases to leak. The explanation \cite{limited} is that the true eigenstate of the Hamiltonian has order unity overlap with the initial state.

Note that stability may not persist in all dimensions; certainly threshold features of the density of states and spectrum are affected by dimension, and the usual intuitions regarding Fermi's Golden rule may again hold sway. In \cite{confine} we made the point that the symmetry breaking of the Jahn-Teller effect makes this a one-dimensional problem, significantly enhancing the possibility of classical breathers. The same is likely to be true quantum mechanically. It may even be that this plays a role in the temperature-dependent decay of the breather (through an effective increase in dimension), as evidenced by the high-temperature disappearance of anomalous decay in doped alkali halides \cite{theoryandfit3}. This issue should be addressable using the path integral method.

In conclusion, we have shown that although quantum tunneling does in principle convert the classical breather into a Bloch state, when the ability to tunnel is removed, an initially localized pulse of energy can be trapped indefnitely as a quantum excitation.

\acknowledgments
This work was supported by NSF grant PHY 00 99471 and Czech grants ME 587 GA AVCR A1010210.

\begin{thebibliography}{16}
\bibitem[{not({\natexlab{a}})}]{note:names}
\bibinfo{note}{Discrete breathers (DB) are a kind of soliton also known as
  ``intrinsic localized modes'' (ILM).}
\end{thebibliography}

\end{document}